\documentclass[fleqn,usenatbib]{mnras}
\usepackage{CJK}
\usepackage{csvsimple}

\usepackage{newtxtext,newtxmath}
\usepackage{CJK}
\usepackage{graphicx}
\usepackage{amsmath}
\usepackage[T1]{fontenc}

\DeclareRobustCommand{\VAN}[3]{#2}
\let\VANthebibliography\thebibliography
\def\thebibliography{\DeclareRobustCommand{\VAN}[3]{##3}\VANthebibliography}

\newcommand\FluxUnit{cm$^{-2}$ s$^{-1}$}

\newcommand\gray{$\gamma$-ray}

\usepackage[dvipsnames]{xcolor}

\title[Gamma-ray Stellar Flares]{Searching For Gamma-ray Emission from Stellar Flares}

\author[Y. Song et al.]{
Yuzhe Song(宋宇哲),$^{1, 2, 3, 4, 5}$\thanks{E-mail: yuzhesong@swin.edu.au}
Timothy A. D. Paglione,$^{3,4,5}$
and Ekaterina Ilin$^{6}$
\\
$^{1}$Center for Astrophysics \& Supercomputing, Swinburne University of Technology, Hawthorn, VIC 3122, Australia\\
$^{2}$OzGrav - ARC Centre for Excellence of Gravitational Wave Discovery, Hawthorn, VIC 3122, Australia \\
$^{3}$Department of Physics, the Graduate Center, City University of New York, 365 Fifth Ave., New York, NY 10016, USA \\
$^{4}$Department of Earth \& Physical Sciences, York
College, City University of New York, 94-20 Guy R. Brewer Blvd., Jamaica, NY 11451, USA \\
$^{5}$Department of Astrophysics, American Museum of
Natural History, Central Park West at 79th Street, New York, NY 10024, USA \\
$^{6}$ASTRON, Netherlands Institute for Radio Astronomy, Oude Hoogeveensedijk 4, Dwingeloo, 7991 PD, The Netherlands \\
}

\date{Accepted 2024 May 24. Received 2024 May 02; in original form 2024 January 23}

\pubyear{2024}

\begin{document}
\begin{CJK*}{UTF8}{gbsn}
\label{firstpage}
\pagerange{\pageref{firstpage}--\pageref{lastpage}}
\maketitle

\begin{abstract}
Flares from magnetically active dwarf stars should produce relativistic particles capable of creating \gray s. So far, the only isolated main sequence star besides the Sun to have been detected in \gray s is TVLM 513-46546. Detecting \gray\ flares from more dwarf stars can improve our understanding of their magnetospheric properties, and could also indicate a diminished likelihood of their planets' habitability. 
In this work, we stack data from the Fermi Gamma-ray Space Telescope during a large number of events identified from optical and X-ray flare surveys. 
We report an upper limit of \gray\ emission from the population of flare stars. Stacking results towards control positions are consistent with a non-detection. We compare these results to observed Solar \gray\ flares and against a model of emission from neutral pion decay. The upper limit is consistent with Solar flares when scaled to the flare energies and distances of the target stars. As with Solar flares, the neutral pion decay mechanism for \gray\ production is also consistent with these results. 

\end{abstract}




\section{Introduction} \label{sec:intro}

The Sun is a \gray\ source both while quiescent \citep{fermi_halo, orlando_strong_2008} and when it flares \citep{ flsf,behind_the_limb, fermi_flare}. The first Fermi-LAT Solar Flare Catalog \citep[FLSF,][]{flsf} provides detailed insights into \gray\ Solar flares, such as their emission mechanism and associations with coronal mass ejections (CMEs). 
Extreme Solar events can severely impact the Earth's atmosphere and even create geomagnetic storms that impact human activities \citep{2022A&A...659A..10V}.
Being more magnetically active than the Sun, red and ultra-cool dwarf stars can generate flares that are orders of magnitude more energetic than Solar flares. The extremely energetic stellar surface activity can lead to proton acceleration events such as  CMEs. Stellar winds, CMEs and accompanying high energy photons can alter the atmosphere and magnetosphere of surrounding planets \citep{ 2022MNRAS.509.5858H, 2021NatAs...5..298C}. 
Some of these stellar flares can be categorised as super- or even mega-flares as they reach output energies over $10^{36}$ erg  s$^{-1}$. Such events are predicted to emit \gray s \citep{OH2018}. These stars also have much higher flare frequencies compared to the Sun. With their large abundance in the Galaxy, and their prevalence for hosting planets, detecting \gray\ flare events from these stars would imply a greatly diminished number of habitable worlds.

Many recent ground- and space-based missions and facilities, searching for transient events at many wavelengths, are able to capture stellar flares and create large flare catalogs. These include TESS \citep{Gunther2019}, Kepler \citep{Hawley2014}, Evryscope \citep{howard2019}, MAXI \citep{maxi_flares}, the Australian Square Kilometer Array Pathfinder \citep{rigney2022}, and the Deeper, Wider, Faster Program \citep{dobie2023}. 
In \citet{song2020}, a $4\sigma$ detection was achieved by phase-folding the lightcurve of TVLM 513-46546, an unusually active, nearby, and rapidly rotating radio dwarf star. However, a residual photon counts stacking search for \gray\ emission from 97 nearby flare stars did not result in a detection. With the advent of extensive flare surveys, we are now able to select a much larger sample with hundreds of stars and thousands of flares to search for stellar \gray s. Further, development of \gray\ stacking techniques using joint likelihoods \citep{principe2021, song2021, paliya2020, ajello2020b, paliya2019} indicates improved sensitivity in mining Fermi-LAT data for sub-threshold signals. In this work, we attempt a stacking method utilising joint likelihoods, combined with a windowing scheme that isolates the \gray\ data around the flare times, to search for \gray s from 1505 flares from 200 flaring dwarf stars.

\section{Observations and Results}
\label{sec:methods}

\subsection{Sample Selection}
\label{subsec:sample}

We utilised optical flare data from the Evryflare \citep{howard2019} and TESS Flare \citep{Gunther2019} surveys. These optical surveys have high cadences of two minutes and cover most of the sky to provide accurate flare times for a very large number of uniformly distributed stars. For example, the Evryscope flare catalog observed  575 flares from 284 stars, and the TESS survey observed 8695 flares from 1228 stars. We also used the all-sky survey of X-ray flares by the Monitor of All-sky X-ray Image (MAXI) project \citep{maxi_flares}, which detected 21 giant flares from 13 active stars. These X-ray flares, while smaller in number compared to the optical flare surveys, are extremely energetic and may therefore be more likely contributors of \gray\ emission within the stacking survey.  
Only stars with Galactic latitude $>20$\degr\ were selected in this study to avoid the complicated \gray\ background caused by the Galactic plane. These surveys were able to identify, in many cases, multiple flares from any individual star during the observation (Table~\ref{tab:flare_surveys}).
Distance cuts were made so that the estimated flare fluxes could be within the range of fluxes that we can conceivably probe with the stacking methods (as high as $\sim 10^{-11}$ \FluxUnit, detailed in \S~\ref{subsec:scaling}). For the Evryflare survey, stars within 50 pc were included, which resulted in 106 stars and 244 flares. For the TESS Flare survey, stars within 25 pc were included, which resulted in 86 stars and 1225 flares. For the MAXI flare survey, 8 stars satisfy the Galactic latitude cut, which included a total of 16 flares. No distance cut is applied to the MAXI flare stars given the low number of stars and extremely high flare energy. The 8 stars have distances ranging from 6.5 pc to 88 pc. 

\begin{table*}
\begin{center}
\caption{Samples Drawn from Each Flare Survey \label{tab:flare_surveys}}

\begin{tabular}{lllllll} 

\hline
 Instrument & Wavelength & Max Dist.  & No. Stars & No. Flares & Reference \\
\hline
  Evryscope & Optical & 50 pc & 106 & 264 & \citet{howard2019}\\
 TESS & Optical & 25 pc & 86 & 1225 & \citet{Gunther2019}\\
  MAXI & X-ray & 88 pc & 8 & 16 & \citet{maxi_flares}\\
\hline
\end{tabular}
\end{center}
\end{table*}

\subsection{Fermi-LAT Data Analysis}
\label{subsec:analysis}
The analysis in this work utilised the third revision of the PASS8 data, P8R3, released on Nov 26, 2018, of $Fermi$-LAT, along with the 10-year source catalog \citep[the 4FGL-DR2,][]{4FGL}, hereafter the 4FGL. The latest Galactic interstellar emission model {\tt gll\_iem\_v07}, and isotropic background model {\tt iso\_P8R3\_SOURCE\_V3\_v1} \citep{abdo2009} were used. Data analysis in this work was performed with {\tt Fermipy} \citep{fermipy} version 1.0.1\footnote{https://fermipy.readthedocs.io/en/latest/} based on the Fermi Science tools Anaconda distribution \citep{anaconda2020} version 2.0.8\footnote{https://fermi.gsfc.nasa.gov/ssc/data/analysis/}. 

For each star, the analysis was performed both on the full mission elapsed time (MET), and within a specified time window (or windows) around the flare(s). The full MET analysis covers MJD = 54678.05 to MJD = 59453.00. The window began just before the peak of the optical or X-ray flare, and extended for many hours beyond the peak, based on the prevalence of Solar \gray\ flares to have such long durations. 
For Evryflare and MAXI targets, we used Fermi-LAT data from 15 minutes before the start of each observed flare, and for 24 hours after the peak time of the flare. For TESS targets, however, in many cases there were a lot of recorded flares within the span of a day. All flares that happened within one day of one another were observed in the same window. A flare window started 15 minutes before the peak time of the first flare, and ended 24 hours after the peak time of the last flare identified in this window. If a target star had more than one flare window, then all the flare windows were combined using the tool {\tt gtselect}. 

Each ROI was a $21.2\degr \times 21.2\degr$ square and centered at the location of a star, which corresponded to a $\sim 15\degr$ radius region of interest (ROI). The data were also filtered using a zenith angle cut of $90\degr$ to avoid bright emission from the Earth. Good time intervals were chosen with conditions {\tt DATA\_QUAL==1 \&\& LAT\_CONFIG==1}. In this analysis, the data were binned uniformly in 37 logarithmically-spaced energy bins, between 300 MeV and 100 GeV. The standard binned likelihood analysis process was performed on each ROI within the observation time periods described above, containing all the observed flare times of each star or the full MET. The star was added as a source with a power law (PL) spectrum to the center of the ROI model. To quantify the significance of any detection, the Test Statistic (TS) of each star was obtained through this process, defined as TS $=2 \log{{\cal L}/{\cal L}_0}$, where ${\cal L}$ is the likelihood from the best-fit model containing the star, and ${\cal L}_0$ is the likelihood of the model for the null hypothesis.

\subsection{Stacking Analysis}
The binned likelihood analysis of Fermi-LAT data returned a model of the ROI which was then used for the stacking analysis. A point source with a PL spectrum was placed at the location of the star, and the flux and spectral index were fixed. Only the background isotropic and Galactic diffuse normalisations were free to be fit by {\tt Fermipy}'s {\tt fit} function. This process returned a log likelihood for the ROI given these spectral parameters for the central source. We repeated this process over a grid of flux and spectral index values. TS maps in flux-index parameter space for each ROI were made by subtracting the log likelihood of the grid point representing the null, which was chosen to be at the lowest flux in the grid ($7.2\times 10^{-12}$ \FluxUnit) and softest spectral index of $-4$, then multiplying this result by 2. The stacked TS map was then made by summing the individual TS maps.

Test sources located at supposedly empty sky locations should be also analysed to serve as a set of controls to compare with the results of the stars. There are two ways of choosing the control comparison: first is to choose an ``off'' time window for each star with the same exposure and perform the same analysis as mentioned above; and second is to choose a random location within each ROI. 
Since we cannot guarantee that these relatively short time windows have the identical exposure in the ``off" time, we perform the control comparison with the latter method. 
Test sources for this analysis were placed at a random location within the ROI five degrees away from the star, within the same flare window, and subject to the same analysis procedures described above. 

\subsection{Results}
\label{subsec:results}
The individual source analysis was first applied to all 200 flare stars using data covering the entireMET. None of the stars have a TS value larger than 25, which traditionally indicates a significant detection. Stacking the fullMET data of these stars also returned results consistent with a non-detection. 

\begin{figure*}
    \centering
    \includegraphics[width=0.45\textwidth]{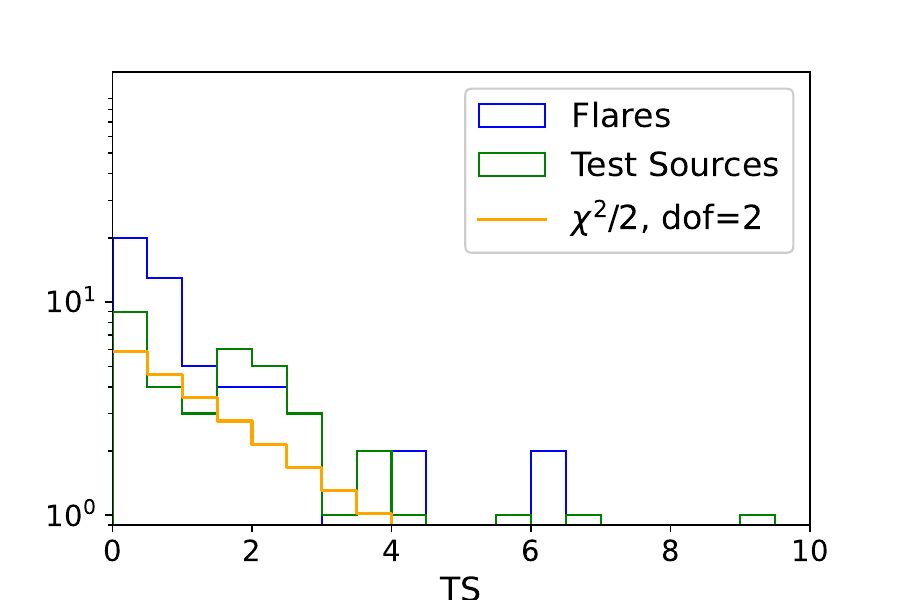}\\
    \includegraphics[width=0.45\textwidth]{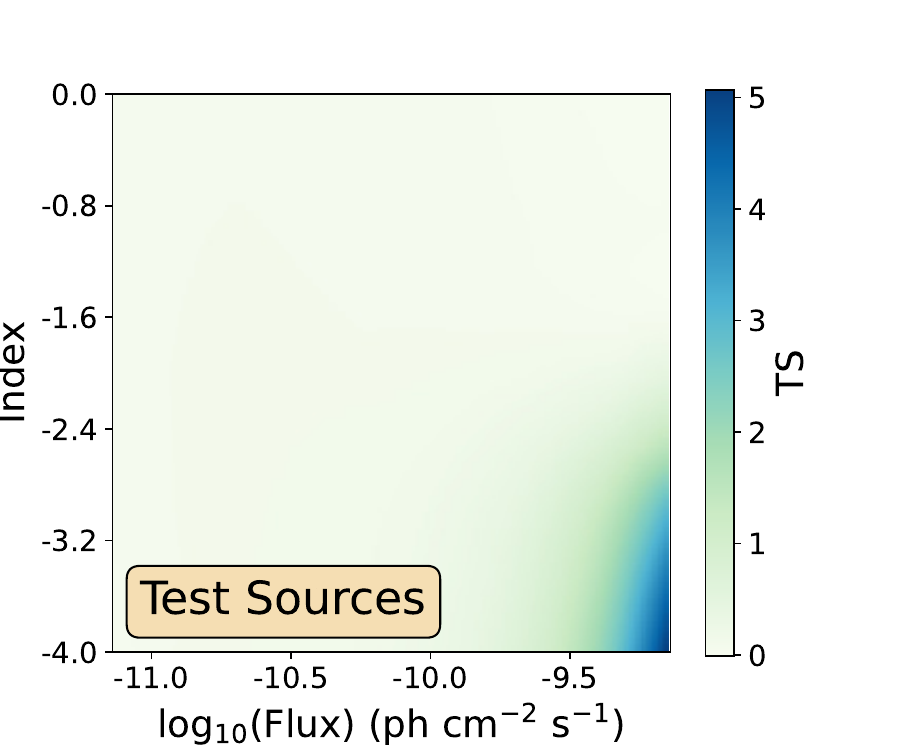}\
    \includegraphics[width=0.45\textwidth]{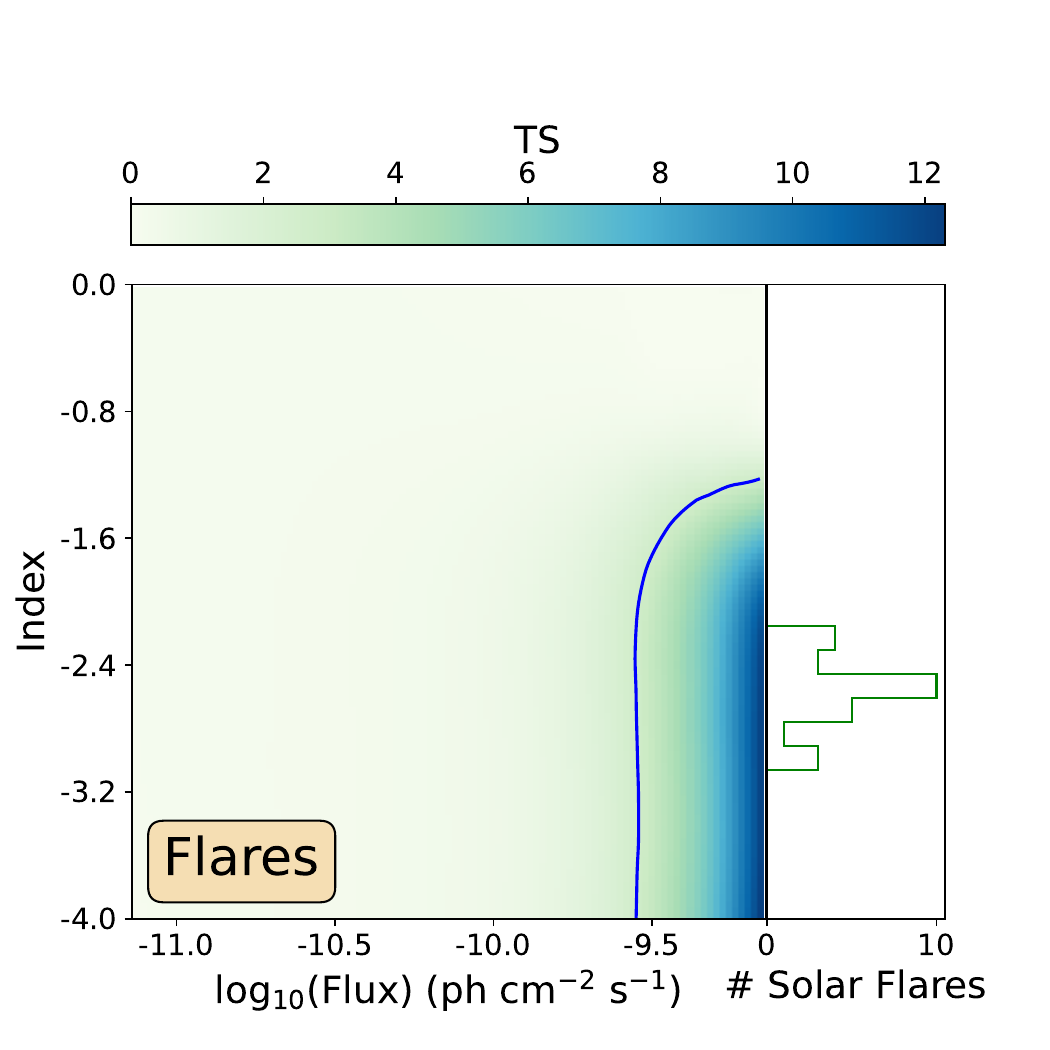}
    
    \caption{Top: Distributions of TS values for: stacked flare windows of each star (blue), test sources (green), and $\chi ^2/2$ distribution with 2 d.o.f. Bottom left: parameter space stacking for the 298 test sources. Bottom right:  parameter space stacking of 53 stars containing 298 observed flares with a TS peak at photon index of $-2.5$ and an upper limit in flux at $1.9 \times 10^{-10}$ \FluxUnit . The blue contour in the figure indicates TS = 2.71, which is traditionally used to indicate the 95\% upper limit. The distribution of the PL index of all 26 FLSF cataloged solar flares with a PL spectral model is displayed on the right.}
    \label{fig:flare_window_stack}
\end{figure*}

When analysing the flare windows, no star had central sources with TS values larger than 25. An overwhelming number of the stars in fact had TS values $\sim 0$. Overall, the TS distributions of both the stars and the test sources are similar to each other as well as to the theoretical null, which is proportional to a $\chi^2$ distribution for two degrees of freedom (Fig~\ref{fig:flare_window_stack}).

We also report the stacked parameter space TS maps of the stars and the controls in the bottom two panels of Fig~\ref{fig:flare_window_stack}. 
Since no individual target sources or control field test sources are detected, which is evident in the TS distributions in the top panel of Fig.~\ref{fig:flare_window_stack}, we only explore the spectral parameter space up to the point source detection sensitivity of the LAT. Any higher, and a source should be detected. 
Out of the 200 stars, 53 converged in the stacking analysis, containing a total of 298 flares. The spectral parameter stack for the flare stars, even though the total TS never reaches 25, peaks at a photon index of $-2.5^{+1}_{-1.5}$. 
The monotonic increase in likelihood with flux is simply indicative of a flux upper limit. 
This result is consistent with the PL spectral index values observed for \gray\ solar flares \citep{flsf}. The stacked TS map for the test sources, however, appears to be a non-detection with its maximum TS value in the corner of the parameter space at the highest flux and the softest spectral index. 
The difference in TS values of the test sources and flares, $\Delta$(TS) $= 7$, also indicates that the existence of flare \gray\ emission is only marginally favored. 
To obtain the 90\% confidence level upper limit of the flux of the stack, the flux of the stack is increased until ${\cal L}_{\rm max} - {\cal L}_{\rm max}|_{90} = 2.71/2$, where ${\cal L}_{\rm max}$ is the maximum likelihood of the null hypothesis, and ${\cal L}_{\rm max}|_{90}$ is the maximum likelihood of the model with increased flux, or $\Delta$(TS) = 2.71 \citep{2012A&A...547A.102H, 1997sda..book.....C}.
Setting the PL index to be $-2.5$, we estimated the upper limit flux of the stacked flares to be $1.9 \times 10^{-10}$ \FluxUnit. Although this flux is about an order of magnitude below the point source sensitivity of the LAT, this analysis remarkably may constrain the PL spectral index of the stellar flares. 

It bears repeating that the TS distributions for the flares and control field test sources are not only indistinguishable from each other, but also from the null. The null distribution is not an isolated peak at TS = 0, but the $\chi^2$, which means it stacks up to a nonzero value. We do indeed observe a non-zero, but insignificant, peak in TS for test sources in the bottom left panel of Fig.~\ref{fig:flare_window_stack}. We also note that the control field stack is spectrally distinguishable from the flares. This behaviour is known from other analyses of control field stacking \citep{song2021,paliya2020} and implies that they trace or resemble the diffuse \gray\ background. 

\section{Discussion} 
\label{sec:discussion}

Compared to \citet{song2020}, the most obvious change made in this work is the much larger sample size. In \citet{song2020}, we examined flare stars that had been detected in radio and/or X-ray surveys, and utilized Fermi-LAT data from the entire MET. Flares from these stars might emit in \gray s, but examining the full MET can dilute the signal and decrease the sensitivity. In this work, we choose optical and X-ray surveys that provide the time of every flare, which allows us to isolate each one individually and avoid signal dilution. More importantly, the analysis methods have been significantly updated and are more sensitive. Rather than stacking residual photon counts, the stacking analysis is now performed on the likelihood profile in spectral parameter space of each flare, which proves to be more sensitive.

\subsection{Flare Frequency Distribution}
\label{subsec:ffd}

Since the Sun is the only star with individual flares observed in \gray s, it is our best template to understand any stacked flare signal in this study. To establish the appropriate context in order to compare stellar and Solar flares, we first examine the flare frequency distributions (FFDs) of all the flares investigated in this work. The FFD describes the rate of flares above a given energy $E$, and typically follows a power law:
\begin{equation}
    f(> E) = \frac{\beta}{\alpha - 1} E^{-\alpha + 1},
\end{equation}

\noindent
where $\alpha$ and $\beta$ are free parameters to be fitted. The power law index $\alpha$ is often used as an indication of the magnetic activity of the stars \citep{1989SoPh..121..375S, 2018AgWM..204..126P}. In examining the FFDs, we can illustrate the differences and similarities between the Solar and stellar flares, and justify the scaling of the Solar flares in the following analysis.

We estimate the total flare energy with  the GOES soft X-ray (SXR) observations of solar flares, catalogued by \citet{sxr_flares}, who provide a detailed list of SXR flares between 1986 - 2020\footnote{https://github.com/nplutino/FlareList}. The flare times in the SXR catalog are matched to the FLSF catalog. All SXR flares that fall between the estimated start and end time of a FLSF flare, are counted towards that FLSF flare. The summed integrated flux of all SXR flares within a FLSF flare, multiplied by $4\pi \textrm{AU}^2$, is the total flare energy. 
The caveat of this estimation is that the total energy of a solar flare released in SXR only serves as a lower limit. A potentially more accurate estimate of flare energy is from the proton energy, given that the flare energy should be 20 times the proton kinetic energy \citep{OH2018}. 
However, it is beyond the scope of this work to correlate SXR luminosity to total flare energy output. For the five FLSF flares that are studied in \citet{2017ApJ...836...17A}, we estimate their flare energies as 20 times the Solar energetic particle energy.  

\begin{figure*}
    \centering
    \includegraphics[width=0.7\textwidth]{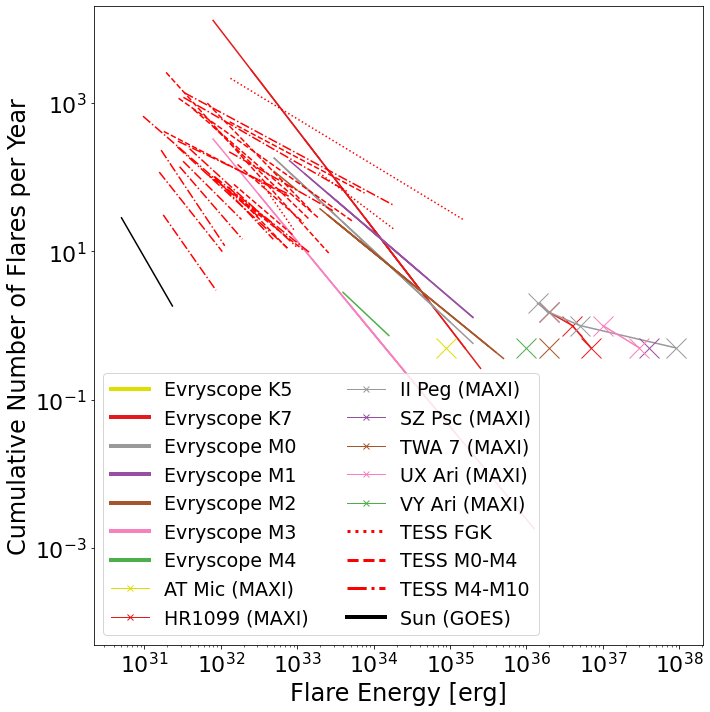}
    \caption{Flare frequency distributions of all flares investigated in this work and the FLSF Solar flares. Flare energies of the Solar flares are estimated as described in the text above using integrated SXR flux. Flare energies of the stellar flares are the values from the respective flare catalog. 
    }
    \label{fig:FFD}
\end{figure*}

The FFDs of the all the flares in this study, as well as all Solar flares in FLFS, are produced using {\tt Altaipony}\footnote{https://altaipony.readthedocs.io/en/latest/\#id10} \citep{Ilin2021altai, 2019A&A...622A.133I, 2016ApJ...829...23D}, in which $\alpha$ and $\beta$ are estimated using an MCMC power law fitting method \citep{2004ApJ...609.1134W}. All the FFDs are presented in Fig~\ref{fig:FFD}. 
Given the same flare energy, the stars in this survey flare far more frequently than the Sun, and their flare energies are also much higher compared even to \gray\ Solar flares. The \gray\ Solar flare FFD is generally very low and steep in comparison.

The vast majority of the flares in the sample can be thought of as extremely high energy versions of solar flares. While the flare energy varies, the power law slope $\alpha$ of the Solar flares FFD are comparable to those of the TESS and Evryscope flares, indicating they have similar physical origins. 
In contrast, the Solar and stellar flares have very different $\alpha$ values compared to the MAXI flares, which is expected as these X-ray megaflares are associated with young stars or RS CVn systems. Regardless, these X-ray megaflares are still not detected in the stack, and contribute to the significance of the stack as much as the rest of the sample. 

\subsection{Comparison With Solar Flares}
\label{subsec:scaling}
Having justified the common origin and scalability of stellar and Solar flares in the previous subsection, we now estimate the expected stellar \gray\ signal based on an examination of the Solar flares in the FLSF observed between 300 MeV and 10 GeV by Fermi-LAT. 
Our stacking results place a sensitive upper limit on the average \gray\ flux from these flares and constrain the PL index of their \gray\ emission. 

As a simple comparison, we first scaled the \gray\ fluxes of the 26 PL solar flares to a distance of 25 pc, the average distance of the 53 stars being stacked. Even the most energetic solar flare only has a \gray\ flux of $1.42 \times 10^{-16}$ \FluxUnit, which is many orders of magnitude below the range of parameter space being examined in this work. These scaling results are plotted as the black data points in Fig.~\ref{fig:models}.

Assuming that flare flux depends linearly on total flare energy (which we substantiate in the next subsection), we can further scale the \gray\ flux of the FLSF solar flares using the estimated Solar flare energy described in \S~\ref{subsec:ffd}, and 
\begin{equation}
    \frac{E_{\rm stellar}}{E_{\rm solar}} \times (\frac{\rm AU}{\rm d})^2,
\end{equation}

\noindent where $E_{\rm stellar} = 2.3 \times 10^{33}$ erg is the median flare energy of the sample stellar flares, $E_{\rm solar}$ is the flare energy of any given Solar flare, and $d = 25$ pc is the average distance to the target stars. The range of these scaled Solar flare \gray\ fluxes is $1 \times 10^{-16}$ to $3 \times 10^{-11}$ \FluxUnit, which is below the LAT sensitivity limit of $\sim 10^{-9}$ \FluxUnit, but overlaps with the fluxes probed by our stacking method.

\subsection{Emission Modeling}
\label{subsec:implication}
In this section, we explore how our sensitive flux upper limit constrains the flare physics for these stars. We use {\tt Naima} \citep{naima, 2014PhRvD..90l3014K}\footnote{https://naima.readthedocs.io/en/latest/index.html}, a Python package that computes radiation from non-thermal particle populations and also does MCMC fitting to spectra. Solar \gray\ flares appear to be well described by the decay of neutral pions which are created when non-thermal protons strike the Solar atmosphere. 
Proton populations with PL spectral indices $\Gamma_{\rm p}$ ranging from $-6$ to $-3.2$ yield spectra consistent with Solar \gray\ flares \citep{flsf}. We use input proton spectra with $\Gamma_{\rm p} = -6$ and $-3$, and three different values for the target density of the stellar atmosphere: $10^{8}$, $10^{10}$, and $10^{12}$ cm$^{-3}$. These reflect the proton densities theoretically estimated for flare stars by \citet{OH2018}, and for the Solar disk model of \citet{1991ApJ...382..652S}, given the average depth for proton absorption. 
For any given combination of $\Gamma_{\rm p}$ and atmospheric density, the proton spectrum normalisation can be determined from the total proton kinetic energy, which is 5\% of the flare energy.

\begin{figure*}
    \centering
    \includegraphics[width=0.7\textwidth]{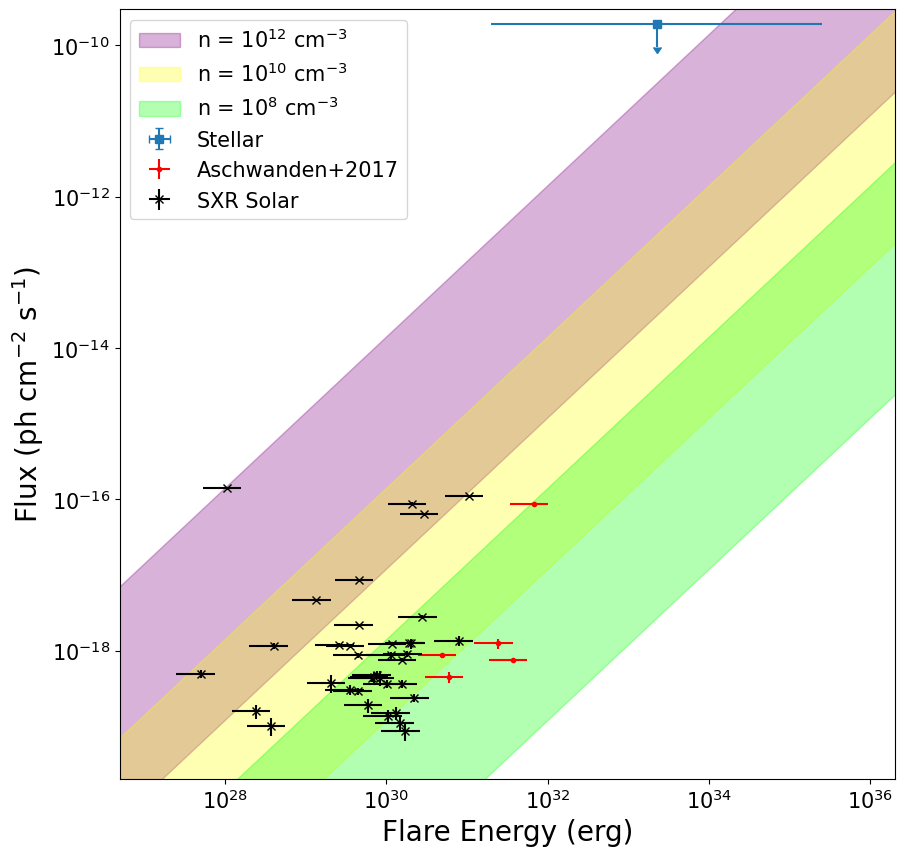}
    \caption{Gamma-ray flux as a function of flare energy for the Solar flares and the stellar flares. Black data points are FLSF Solar flares plotted with SXR solar flare energy estimation; red data points are FLSF Solar flares that have SEP energy estimates from  \citet{2017ApJ...836...17A}. The blue upper limit is the stellar \gray\ flux. Its flare energy value and uncertainty are the average and range of all the flare energies used in the stack. The overlapping green, yellow and purple shaded areas represent the results from the pion decay modeling implemented with {\tt Naima}. Each shaded region represents a different target proton density, and their vertical limits depend on proton index: $-3$, bounded from above, and $-6$, bounded from below. }
    \label{fig:models}
\end{figure*}

The model results for each proton density are plotted as the overlapping shaded areas in Fig~\ref{fig:models}. The lower and upper boundaries of each area are for $\Gamma_{\rm p} = -6$ and $-3$, respectively. The \gray\ flux of Solar flares as a function of  their SXR flare energy estimation, and the upper limit from the stacking results are also plotted. The pion decay models are consistent with both the Solar flares as well as the upper limit from the stellar flare stack. This further indicates that emission mechanism of stellar superflares is likely similar to that of Solar flares. Additionally, the results shown in Fig.~\ref{fig:models} indicate that the energy conversion from flare to proton kinetic energy should not be more efficient than 5\%. If more flare energy were converted to proton energy, creating the same level of \gray\ flux would require a less energetic flare, which would start to contradict the upper limit. A recent study by \citet{2023ApJ...944..192K} indicated that no more than 0.1\% of total flare energy output is converted into non-thermal protons, which is also consistent with our upper limit.

We note that the predicted photon spectral indices for Solar flares, using a power-law with exponential cutoff (PLEC) model, ranges from 3.5 to 4.5 \citep{2018ApJ...864..148K}. Due to the low detection significance of our results, we do not use PLEC models, only power-law models. For this reason, we cannot directly compare the upper limit to these model predictions. However, this range agrees with those \gray\ Solar flares from the FLSF catalog modeled with the PLEC spectrum.

\subsection{Prospect of TeV Observations}

\citet{OH2018} suggested that TeV emission should be present from stellar flares. Very high energy observatories, such as SHALON, recently claimed detection of TeV emission from the direction from M dwarfs \citep{2019AdSpR..64.2585S}. The Cherenkov Telescope Array\footnote{https://www.cta-observatory.org/} (CTA) should be sensitive enough to observe TeV stellar flares, and in fact, modeling from \citet{OH2018} suggests superflares from DG CVn will be detectable by CTA. 
If Target of Opportunity (ToO) observation is adopted by the CTA consortium as part of the observing plans, it can be taken advantage of to detect TeV flares. Combined with the large number of flares anticipated \citep{2024arXiv240206002C, 2009AJ....138..633K} from the Vera C. Rubin Observatory \citep{2019ApJ...873..111I}, and broker software such as Fink\footnote{https://fink-broker.org/} \citep{2021MNRAS.501.3272M}, CTA can quickly slew towards the flaring star for followup TeV observation. 
Targets for the ToO observations can be triggered follow-ups from ground based all-sky monitoring missions. Evryscope introduces a pipeline for low-latency transient detection which is suitable for detecting superflares \citep{2023ApJS..265...63C}. These triggered events could potentially be used for low-latency follow-up with CTA.
Possibly included in these triggered follow-ups, TRAPPIST-1 would be an interesting source to focus on. At a distance of 12 pc and predicted to have $4^{+1.9}_{-0.2}$ superflares per year \citep{2020ApJ...900...27G}, it should be at least as detectable in TeV as DG CVn. Detecting GeV and TeV \gray\ emission around this planet-hosting star can help further understand habitability of exoplanets. Additionally, recent JWST observations of TRAPPIST-1 of transits during flares \citep{2023ApJ...959...64H} could be useful for atmospheric characterization efforts. Multi-wavelength observations of TRAPPIST-1 planetary transits during flares could potentially be helpful towards these efforts.

\section{Conclusion}
In this work, we used sensitive stacking methods to search for any potential \gray\ emission associated with energetic stellar flares. Stacking the LAT data using the fullMET shows no detection, while gating the data around the flare times returns a sensitive upper limit of the flare \gray\ emission. The same analysis on empty test locations as a control returns a null result. Modeling this upper limit with {\tt Naima} and comparing it with Solar \gray\ flares indicates that the common emission mechanism is likely neutral pion decay generated in the stellar atmosphere during flare events. 
To remain consistent with the stellar flare upper limit, the proton acceleration efficiency should not exceed 5\%.

\section*{acknowledgements}
This work was supported in part by NSF grant AST-1831412 and the PSC-CUNY award \# 63785-00 51. The first author is also supported by the Australian Research Council Centre of Excellence for Gravitational Wave Discovery (OzGrav), through project number CE170100004. This project made use of computational systems and network services at the American Museum of Natural History supported by the National Science Foundation via Campus Cyberinfrastructure Grant Awards \# 1827153 (CC$*$ Networking Infrastructure: High Performance Research Data Infrastructure at the American Museum of Natural History).

\section*{Data Availability}
Due to the size of output files for {\it Fermi}-LAT data analysis, the authors can share configuration files used to process all LAT data.
Post processing scripts, including the stacking analysis, will be hosted on Github as well. Before made available on Github, they will be available upon request provide this work is properly cited. 

\bibliographystyle{mnras}
\bibliography{stack_flares}

\bsp	
\label{lastpage}
\end{CJK*}
\end{document}